\documentstyle[12pt,epsf]{article}

\large
\newcommand{\be}{\begin{eqnarray}}
\newcommand{\ee}{\end{eqnarray}}

\begin{document}
\setlength{\baselineskip}{23pt}
\setlength{\baselineskip}{27pt}
\pagestyle{empty}
\renewcommand{\thefootnote}{\fnsymbol{footnote}}
\centerline{\bf\LARGE  Bose Condensate}
\centerline{\bf\LARGE made of Atoms with Attractive Interaction }
\centerline{\bf\LARGE is Metastable}
\vskip 1cm
\centerline{\bf E.V.~Shuryak}
\vskip 1cm
\centerline{\it Department of Physics}
\centerline{\it State University of New York at Stony Brook}
\centerline{\it Stony Brook, New York 11794, USA}
\vskip 1cm

\setlength{\baselineskip}{16pt}
\centerline{\bf Abstract}
  Recent experiments with trapped cooled atoms have produced evidences for
 Bose-Einstein condensation. Among the atoms used are $^7Li$ ones,
with attractive low energy
 interaction. Stability of their
condensate is studied, with a conclusion that
for a number of atoms below
the critical value there exist a metastable minimum, separated from collapse
by a potential barrier.
The lifetime due to tunneling is
estimated:  it is unlikely to happen in the present experiment.
Although in experimental conditions most of the atoms are not
in the condensate,
we argue that they only slightly affect the stability
conditions for the condensate evaluated at T=0.

\vfill
\begin{flushleft}
SUNY-NTG-95-49\\
November 1995
\end{flushleft}
\eject
\newpage
\setlength{\baselineskip}{23pt}
\pagestyle{plain}
\renewcommand{\thefootnote}{\arabic{footnote}}
\setcounter{footnote}{0}
\setcounter{page}{1}

   Bose-Einstein condensation is a generic quantum phase transition
discussed in most textbooks on quantum
statistical mechanics. Recently
 several experiments \cite{bose_exp,Li_exp}
with atoms trapped in magnetic traps and cooled to temperatures as low as
$T\sim 100 nK$ have been able to observe it.
It is very significant step forward, because unlike liquid He,
it is a condesate in the low density
regime.

   Among those experiments we concentrate on that performed in
Rice University \cite{Li_exp} with $^7Li$ atoms. The reason is
the $negative$ sign of the
 scattering length of these atoms in the corresponding spin state,
$a=-27.3\pm .8 a_0$ \cite{scatt_len} ($a_0$ is  Bohr radius). Since
classic papers on the interacting Bose gas
(e.g.\cite{LHY}) it is known  that macroscopic system of bosons
 with attraction is
{\it unstable against collapse}.
 However, it was argued many times that the situation may be
 different for a $finite$ number of atoms N trapped in a
$finite$ volume for $finite$ time.  In this paper we discuss the
conditions under which
a $metastable$
Bose-condensed state  exists.

   We start with the T=0 case, in which all atoms are in the same
quantum state, and present
a suggestive dimensional argument why it $may$ exist at all.
It is based
on simple variational approach.
Let $\psi(x)$ be the  ground state
wave function (normalized  by $\int d^Dx|\psi(x)|^2=1$).
In the low energy approximation one can describe interatomic interaction by
the well-known Gross-
Pitaevskii (or the $non-linear$ Schreodinger) equation \cite{GP} which
follows from the Hamiltonian
\be
H=N \int d^Dx [{\hbar^2 \over 2m} |\partial_x \psi(x)|^2 + V(x) |\psi(x)|^2
+ N U_0 |\psi(x)|^4/2]
\ee
where V(x) is the trapping potential while
$U_0=4\pi\hbar^2 a/m$ is the Fermi pseudopotential for point-like interactions.

  Suppose the atoms are in a state with a wave function with
some (variable) size R, $\psi=f(r/R)$. Their kinetic energy
is $O(R^{-2})$, the potential one is $O(R^{2})$, while
the interaction term is
$O(R^{-D})$. So, for the $one-dimensional$ case (D=1) the kinetic energy
 dominates at small R, always
allowing for stable solution. However it is no longer so for more dimensions.
In the two-dimensional case
kinetic and non-linear terms may balance each other, while for the
 D=3 case
the interaction term takes over at small sizes.
  If it is negative, the collapse is
inevitable. Nevertheless,
 the effective potential (which is obtained by the substitution
of the trial function into the Hamiltonian)
 \be
V_{eff}(R)=<H>=C_1/R^2+C_2 R^2
-C_3/R^3 \ee
 may have a  minimum, provided $C_3$ (proportional to the number of atoms),
 is not too large.
However, in order to confirm its existence,
one should be sure that the barrier is there in the whole functional space
of possible quantum states.

In what follows we assume for simplicity that the trap is the 3-dimensional
isotropic oscillator, and  introduce the
 dimensionalless
coordinates  $r=x(m\omega/\hbar )^{1/2}$, measuring time  in $\omega^{-1}$
units, etc. This leads to
\be
H/(\hbar  \omega N)= \int d^Dr [|\partial_r \psi(r)|^2/2 + (r^2/2) |\psi(r)|^2
- (N/N_0) |\psi(r)|^4/2]
\ee
where $N_0=(\hbar  /m\omega)^{1/2}/(2\pi |a|)$ is some characteristic number of
atoms. Since  the experimental trap oscillator unit is
about 3$\mu$m  and it is much
larger than the scattering length, there appears a large number $N_0 >>1$ .
The corresponding  equations, both static and time-dependent
ones,
were studied in some details in \cite{NonlinSCH}. Although these authors
were mostly interested in
repulsive interaction, they have also obtained a very important result
for attractive case, namely the
existence of a static solution for attractive for $N$ below
the critical point
\be
N_{crit}\approx 3.6 N_0 \ee
(about 1000 atoms in the experimental conditions).

   Solution of Schreodinger equation is an extremum of the Hamiltonian:
but it may be either an unstable one (a saddle point) or a local minimum.
In order to find out what is the case one
should look at the surrounding states. Again,
it can be most easily be done by
extensive variational studies of the
Hamiltonian (2). We have used a number of trial functions, of various shapes
(e.g. a sum of several
Gaussians with variable amplitudes and widths).
One obvious approach is to minimize
the expectation
value of H by the steepest descent. (Such ``relaxation"
approach has obvious advantages over studies of time-dependent
Schreodinger equation \cite{NonlinSCH}, which conserves energy
and thus makes it difficult to penetrate into a local minima.)
We have then basically
reproduced
the solution discussed in \cite{NonlinSCH}:
the corresponding  wave function
dependence on
the particle number N is shown in Fig.1(a). Note that
 although density at the origin grows by a significant factor, nothing
in the wave function itself suggests that, as one approaches
the
critical point,
  a qualitative change
should occur.

  The issue of stability was addressed in two steps. First, we have
added to the wave function various $small$ perturbations of $random$ shapes,
and evaluate their energies.
While doing so it  became clear that
the solution $is$ indeed a minimum
(in the indicated N interval). Furthermore, it was found that  if one starts
 minimization from $any$ perturbed trial functions, for $N<N_{crit}$ it
converges back to the original state while
for $N>N_{crit}$ energy minimization
leads to a collapse,  a compressing wave package with
decreasing $negative$ energy.

   The second step was a
  more systematic study of the instability threshold and the barrier
around the metastable state. To this
end, we have written the wave function as the metastable one
$\psi_{ms} $  plus a
perturbation of $finite$ magnitude, e.g.
\be
\psi(x) \sim [\psi_{ms} + C exp(-x^2/R^2)]
\ee
where C,R are variable parameters (a common multiplier is determined
by normalization to 1). A typical result for $N=3 N_0$ is shown in Fig.1
(b): it display energy versus C for variable size R. The barrier is observed
in all cases.  Furthermore, we have verified
 that at the critical point (4)
the simplest ``opening of the pocket" scenario does indeed take place:
for larger N the system may roll down $classically$ into a collapse.

   The next issue is a $lifetime$ of the metastable state.
 We have seen that finite
increase in density at the origin leads to an instability.
Such perturbation may appear spontaneously, as a
result of quantum fluctuation.
Usually one solves such problems semiclassically, by finding a
solution of time-dependent Schreodinger equation with inverted potential
. This solution should start (and end) with the minimum $\psi_
{ms}$, and describe a ``bounce" from  the state $\psi_{outside}$
on the outside part
of the barrier, with the  energy identical to the original one.
 The
penetration probability is then obtained
in terms of the action for that solution
$\sim exp(-S[\psi])$.
 Such approach is known in many similar
cases, see e.g. Coleman's solution for
``the fate of a false vacuum" \cite{Coleman} of the $\phi^4$ relativistic
field theory. Unfortunately, in the present non-
relativistic problem there is no  symmetry between time and space coordinate,
which simplify it so that analytic solution becomes possible.

 Instead of looking for numerical solution, we
suggest another
 estimate for the collapse rate based on $projection$ of two  wave
functions in question, $\psi_{ms}(x)$ and
$\psi_{outside}(x)$,  the other end of the tunnel:
\be
P_{collapse}/\omega \sim  |<\psi_{outside}|\psi_{ms}>|^{2N}
\ee
We do not know which $\psi_{outside}(x)$ is connected
with $\psi_{ms}(x)$ by the classical path (of $minimal$ action),
 but in fact the transition may
happen along any path (and actually in (6) one should sum over all of them).
We have evaluated
the projection probability for a number of $\psi_{outside}(x)$,
which can be found for any direction in the functional space
(for example, for any barrier curve shown in Fig.1(b)). Fortunately, we found
that the projection is in fact
rather insensitive to the
details of the density fluctuation itself, such as
its size R and shape. (It is hardly
worth presenting those details
because variations are  small and also because the
 estimates under consideration are qualitative  anyway.). A typical
results are
 shown in Fig.1(c): they are then  translated into
$P_{collapse}/\omega\sim exp[-.57 \times (N_{crit}-N)]$.
 Although for one or few
 particles it would not be an
improbable fluctuation, for the condensate made of $N\sim 1000$ particles
(which presumably occured in experiment)
the tunneling is strongly suppressed. In experimental
conditions the observation time is about $10^5$ or so oscillation
periods. We therefore conclude that in that experiment the tunneling
 is very improbable, except maybe very close to the critical
point.

  Let us now proceed to the
non-zero temperature case. Although the temperature T is not reliably
measured in the experiment under consideration, it is suggested to be about
$T\sim 150 nK$.
 As $T>>\hbar \omega\sim
5 nK$, for non-condensate particles one can use  classical
 approximations. Furthermore, we ignore self-interaction and set
chemical potential to zero,
looking for  the $maximal$ non-condensate particle density
\be
\rho_{nc}(x)= \int ({dp\over 2\pi \hbar})^D {1 \over exp(p^2/2mT+V/T)-1}
\ee
This leads to the total number of non-condensate particles $N_{nc}=(T/\hbar
\omega)^3 \zeta(3)  $, or
about $\sim 2\times 10^4$ atoms for T mentioned above.
It is qualitatively consistent with the total
number of particles actually observed. This estimate,
together with upper limits on the number of condensate
particles (4),  leads to
conclusion that only several percent of the atoms can belong to the condensate.
(This conclusion is of course a subject to direct experimental test.)

  The last point: can the non-condensate particles significantly
affect the stability condition
discussed above for T=0? Let us compare the total density at the origin
of both components, the non-condensate ones (7)
$\rho_{nc}(0)= ({m T \over 2\pi \hbar^2})^{3/2} \zeta(3/2) $
 and $\rho_{c}=N|\psi_{ms}(0)|^2$. Putting in numbers for the
experimental conditions one concludes, that
the former one is only few percent of the latter. Therefore, the
majority of particles are too far from the origin, and thus the
stability limit at the observed T may only be slightly reduced
compared to the T=0 one (4).

   We have not studied the collapse in this work, which is a formidable task
by itself. Let us only add
few comments about it. First of all, it  certainly is there in
the low energy approximation only, in which the interaction
is assumed to be represented by the scattering length.
The next term in the Hamiltonian $\sim (\partial\psi)^2 (\psi^2)$
(and others indicated
repulsion of atoms at small distances) should stop the
collapse and result in a condensed ground state, solid or liquid one.
The collapse itself deserves separate studies, because it is
related to many other phenomena in physics, from sonoluminescence to
supernovae, which are not yet well understood.  The
energy per particle released in the collapse
 is very large compared to the intial
T, which should lead to strong reheating of the system.
For Li atoms used in experiment \cite{Li_exp}, the  binary potential
in appropriate state
is rather
small (by normal atomic standards): but it is still deep enough to
cause significant reheating, up to temperatures of
the order of at least degrees K (compared with 150 $nK$ at the beginning).
 A ``mini-supernova" event would include
tiny ``hot" cluster (which is no longer a Bose condensate, but liquid or solid)
which eventually would blow up the non-condensate cloud.

In summary: we have established that N trapped atoms with attractive
interaction posess a metastable state, surrounded by a barrier,
provided $N<N_{crit}=3.6 N_0  $. We have studied the barrier
and estimated the tunneling probability. It seems very unlikely that
tunneling could occur during the Rice group experiment: but it can be
studied in future experiments.
We also argue that under experimental conditions
most particles are $not$ in Bose condensate,
but the non-condensate component does not significantly
affect the stability limits.
 Finally, we speculate that collapse
should be explosive, leading to strong reheating of the small claster
of ordinary condense matter.

{\it Acknowledgements} This work is partly supported by the DOE grant
-FG02-88ER40388.
 I  thank P.Koch who initiated my interest in the problem,
T.Bergeman and C.N.Yang for discussion, and J.Marburger for explanation
of some regimes in two-dimensional case known in laser physics.

Note added: when the paper was completed we have learned about a
relevant recent preprint \cite{DS}, in which T=0 case for $anisotropic$
trap was considered and  more accurate
critical number of Li atoms for
the specific trap used in experiment is $N_{crit}= 1400$. Furthermore, they
shown that states with non-zero $vorticity$ are much more stable
because their density at the origin is depleted: those
states are other candidates for
the metastable states seen in experiment, and diffraction data which are
sensitive to condesate shape may in principle separate those from the
lowest state discussed above.

\newpage
{\Large\bf figure captions}\\ \\ \\
(a) The  wave function $\psi_(x)$ of the metastable state, in
oscillator units defined in the text. Different curves
correspond to 4 values of the  number
of atoms N given in the figure (from upper to lower one, at x=0).
(b) Average energy for different trial functions (5), plotted as a function
of
 the wave function at the origin $\psi(0)$. All curves are for $N/N_0=3.$
, for different spacial size of the perturbation: $1/R^2=
10,5,4,3$ for open squares, closed squares, dots and triangles.
All of them display existence of a barrier, separating the metastable state.
 (c) The projection of the state $\psi_{outside}(x)$
(which is  outside the  boundary of stability but have  the same energy as the
the metastable state $\psi_{ms}(x)$ versus the number of atoms N. At the
critical point, the barrier disappear and here the projection becomes 1.
\pagestyle{empty}
\newpage
\begin{figure}
\epsffile{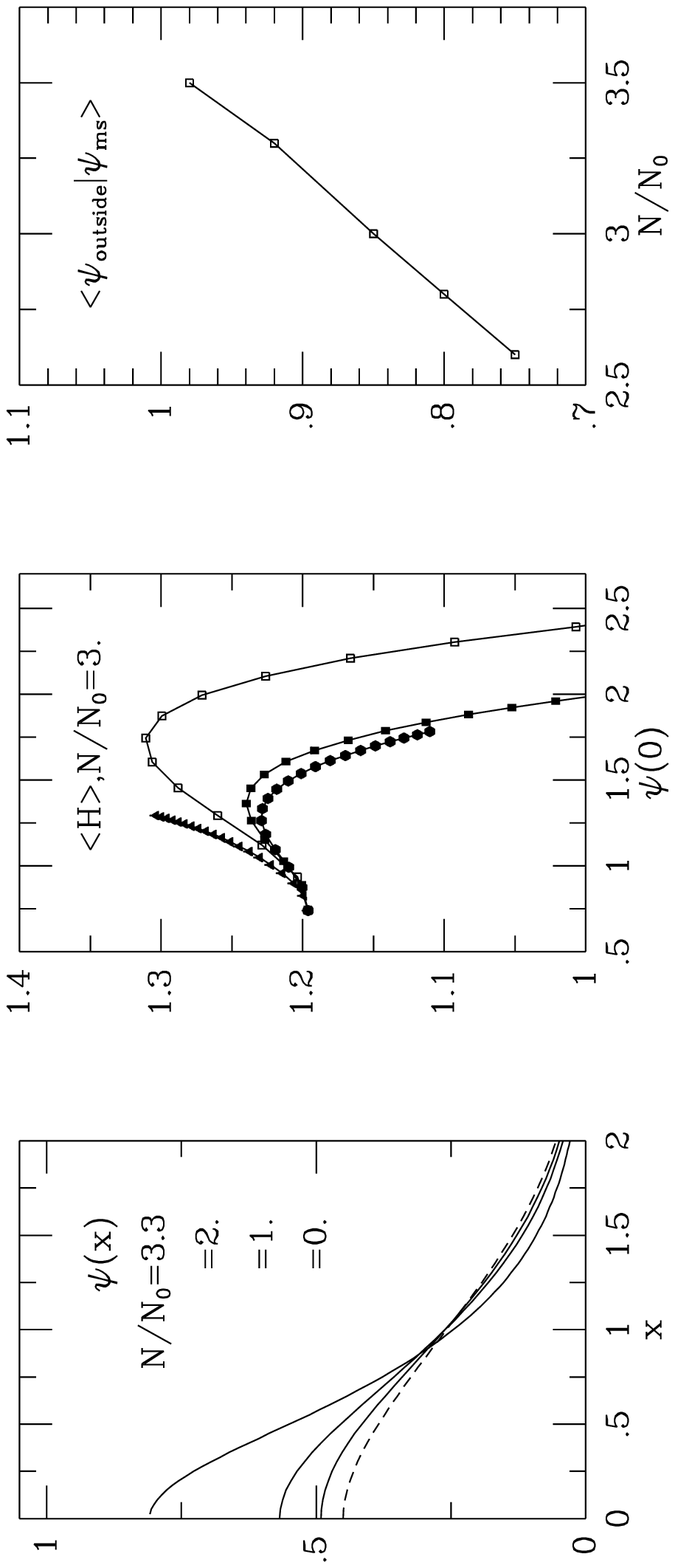}
\end{figure}
\vfill

\end{document}